\documentclass{PoS}

\title{The On-Site Analysis of the Cherenkov Telescope Array}

\ShortTitle{The On-Site Analysis of the Cherenkov Telescope Array}

\author{\speaker{Andrea Bulgarelli}$^{a}$, Valentina Fioretti$^{a}$, Andrea Zoli$^{a}$, Alessio Aboudan$^{a,b}$, Juan Jos\'e Rodr\'{\i}guez-V\'azquez$^{c}$, Giovanni De Cesare$^{a}$, Adriano De Rosa$^{a}$, Gernot Maier$^{d}$, Etienne Lyard$^{e}$, Denis Bastieri$^{f}$, Saverio Lombardi$^{g}$, Gino Tosti$^{h}$, Sonia Bergamaschi$^{i}$, Domenico Beneventano$^{i}$, Giovanni Lamanna$^{j}$, Jean Jacquemier$^{j}$, Karl Kosack$^{k}$, Lucio Angelo Antonelli$^{g}$, Catherine Boisson$^{l}$, Jerzy Borkowski$^{m}$, Sara Buson$^{f}$, Alessandro Carosi$^{g}$, Vito Conforti$^{a}$, Pep Colom\'e$^{n}$, Raquel de los Reyes$^{o}$, Jon Dumm$^{p}$, Phil Evans$^{p}$, Lucy Fortson$^{p}$, Matthias Fuessling$^{r}$, Diego Gotz$^{k}$, Ricardo Graciani$^{t}$, Fulvio Gianotti$^{a}$, Paola Grandi$^{a}$, Jim Hinton$^{o}$, Brian Humensky$^{w}$, Susumu Inoue$^{s}$, J\"urgen Kn\"odlseder$^{x}$, Thierry Le Flour$^{j}$, Rico Lindemann$^{\circ}$,  Giuseppe Malaguti$^{a}$, Sera Markoff$^{u}$, Martino Marisaldi$^{a}$, Nadine Neyroud$^{j}$, Luciano Nicastro$^{a}$, Stefan Ohm$^{s}$, Julian Osborne$^{p}$, Igor Oya$^{\circ}$, Jerome Rodriguez$^{k}$, Simon Rosen$^{p}$, Marc Ribo$^{v}$, Alessandro Tacchini$^{a}$, Fabian Sch\"ussler$^{\star}$,  Thierry Stolarczyk$^{k}$, Eleonora Torresi$^{a}$, Vincenzo Testa$^{g}$, Peter Wegner$^{\circ}$, Amanda Weinstein$^{z}$ for the CTA Consortium\footnote{Full consortium author list at http://cta-observatory.org}
\\
     \llap{$^a$}INAF/IASF Bologna, Bologna, Italy\\
     \llap{$^b$}CISAS, University of Padova, Padova, Italy\\
     \llap{$^c$}Centro de Investigaciones Energ\'eticas, Medioambientales y Tecnologicas, Madrid, Spain\\
     \llap{$^d$}Deutsches Elektronen-Synchrotron, Platanenallee 6, Zeuthen, Germany\\
     \llap{$^e$}University of Geneva,  ISDC Data Center for Astrophysics, Chemin dÕEcogia 16, 1290 Versoix, Switzerland\\
     \llap{$^f$}Dipartimento di Fisica, Universitˆ degli Studi di Padova, Italy\\
     \llap{$^g$}INAF - Osservatorio Astronomico di Roma, Rome, Italy\\
     \llap{$^h$}University of Perugia, Perugia, Italy\\
     \llap{$^i$}University of Modena and Reggio Emilia, Department of Engineering, Modena (Italy)\\
     \llap{$^j$}Laboratoire dÕAnnecy-le-Vieux de Physique des Particules, UniversitŽ de Savoie, CNRS/IN2P3, France\\
     \llap{$^k$}Irfu/SAp, CEA Saclay, Orme des merisiers, b\^at. 709, 91191 Gif-sur-Yvette Cedex, France\\
     \llap{$^l$} Observatoire de Paris, LUTH, CNRS, Universit\'e Paris Diderot, France\\
     \llap{$^m$} Copernicus Astronomical Center, Polish Academy of Sciences, Poland\\
     \llap{$^n$}Institut de Ci\`encies de l'Espai (ICE-CSIC) \& Institut d'Estudis Espacials de Catalunya (IEEC), Spain\\
     \llap{$^o$}Max-Planck-Institut f\"ur Kernphysik, D 69029 Heidelberg, Germany\\
     \llap{$^p$}University of Iowa, Department of Physics and Astronomy, USA\\
     \llap{$^q$}Dept. of Physics and Astronomy, University of Leicester, United Kingdom\\
     \llap{$^r$}Institut fr Physik und Astronomie, UniversitŠt Potsdam, Karl-Liebknecht-Strasse 24/25, D 14476 Potsdam, Germany
     \llap{$^s$}Institute for Cosmic Ray Research, University of Tokyo, Tokio, Japan\\
     \llap{$^t$}Departament dÕAstronomia i Meteorologia, Institut de Cie`ncies del Cosmos, Universitat de Barcelona, Spain
     \llap{$^u$}API/GRAPPA, University of Amsterdam, Amsterdam, The Netherlands\\
     \llap{$^v$}Institut de Ci\`ences del Cosmos, Universitat de Barcelona, Barcelona, Spain\\
     \llap{$^w$}Department of Physics and Astronomy, Barnard College; Department of Physics, Columbia University, USA\\
     \llap{$^x$}CNRS/IRAP, 9 Av. colonel Roche, BP 44346, F-31028 Toulouse Cedex 4, France\\
     \llap{$^z$}Department of Physics and Astronomy, Iowa State University, Ames, IA 50011, USA\\
     \llap{$^{\circ}$}Deutsches Elektronen-Synchrotron (DESY), Platanenallee 6, D-15738 Zeuthen, Germany\\
     \llap{$^{\star}$}Irfu/SPP, CEA Saclay, b\^at. 141, 91191 Gif-sur-Yvette, France\\
     E-mail:  \email{andrea.bulgarelli@inaf.it}}
        
\abstract{The Cherenkov Telescope Array (CTA) observatory will be one of the largest ground-based very-high-energy gamma-ray observatories.
 
The On-Site Analysis will be the first CTA scientific analysis of data acquired from the array of telescopes, in both northern and southern sites. The On-Site Analysis will have two pipelines:  the Level-A pipeline (also known as Real-Time Analysis, RTA) and the level-B one. The RTA performs data quality monitoring and must be able to issue automated alerts on variable and transient astrophysical sources within 30 seconds from the last acquired Cherenkov event that contributes to the alert, with a sensitivity not worse than the  one achieved by the final pipeline by more than a factor of 3. The Level-B Analysis has a better sensitivity (not be worse than the final one by a factor of 2) and the results should be available within 10 hours from the acquisition of the data: for this reason this analysis could be performed at the end of an observation or next morning.
 
The latency (in particular for the RTA) and the sensitivity requirements are challenging because of the large data rate, a few GByte/s. The remote connection to the CTA candidate site with a rather limited network bandwidth makes the issue of the exported data size extremely critical and prevents any kind of processing in real-time of the data outside the site of the telescopes. For these reasons the analysis will be performed on-site with infrastructures co-located with the telescopes, with limited electrical power availability and with a reduced possibility of human intervention. This means, for example, that the on-site hardware infrastructure should have low-power consumption. A substantial effort towards the optimization of high-throughput computing service is envisioned to provide hardware and software solutions with high-throughput, low-power consumption  at a low-cost.
 
This contribution provides a summary of the design of the on-site analysis and reports some prototyping activities.
}

\FullConference{The 34th International Cosmic Ray Conference,\\
		30 July- 6 August, 2015\\
		The Hague, The Netherlands}

\begin{document}

\section{Introduction}

The Cherenkov Telescope Array (CTA) \cite{bib:CTA} will be  the biggest ground-based very-high-energy (VHE) $\gamma$-ray observatory of the future. At the time of writing, the international CTA consortium counts more than 1000 scientists from 28 countries. CTA will consist of two arrays of tens of telescopes: one in the southern hemisphere, to observe the wealth of sources in the central region of our Galaxy, and one in the north, primarily devoted to the study of the Extragalactic sky, providing for example access to Active Galactic Nuclei (AGN) and galaxies at cosmological distances., and star formation and evolution. To accomplish the science goals three different telescope types will be required: a small number of Large Size Telescopes (LST) for the lowest energies (20 GeV - 1 TeV), 20-30 Medium Size Telescopes (MST)  for the 100 GeV - 10 TeV energy domain, and, in the southern hemisphere, tens of Small Size Telescopes (SST) for the highest energies (few TeV - beyond 100 TeV). Thanks to this configuration CTA will achieve a factor of 10 improvement in sensitivity from some tens of GeV to beyond 100 TeV with respect to existing Cherenkov observatories (HESS\cite{bib:HESS}, MAGIC\cite{bib:MAGIC} and VERITAS\cite{bib:VERITAS}). With a total collection area of $\sim 10$ km$^2$, and the improved $<0.1^{\circ}$ angular resolution, CTA will open a large discovery potential in astrophysics and fundamental physics.

Thanks to the large number of individual telescopes, CTA can operate in a wide range of configurations, enabling observations with multiple sub-arrays targeting and simultaneous monitoring of different objects or energy ranges. With its large detection area, CTA will resolve flaring and time-variable emission on sub-minute time scales.  CTA will be 10000 times more sensitive to flares than Fermi at 25 GeV, and combining both sites will achieve full sky coverage. Current CTA simulations show that extreme AGN outbursts, which in the past have reached flux levels ten times the Crab flux, could be studied with a time resolution of seconds, under virtually background-free conditions \cite{bib:CTA}. In addition, some studies \cite{bib:GW}  indicate that the CTA will be capable of following up Gravitational Waves event candidates over the required large sky area (also 1000 deg$^2$) with sufficient sensitivity to detect short gamma-ray bursts, which are thought to originate from compact binary mergers. 

To maximize the science return on time-variable and transient phenomena from astrophysical sources, the CTA Observatory will be capable of both receiving alerts from external observatories and  issuing alerts  by CTA itself during the observations (e.g. through Virtual Observatory notices and alerts), changing the target (i.e. transition from data-taking on one target to data-taking on another target) anywhere in the observable sky within 90 s. This fast reaction to unexpected transient  events is a crucial part of the CTA observatory to better understand the origin of their emission. Thanks to this fast reaction CTA will be capable of following the astrophysical phenomena in real-time and issuing alerts to other facilities, enabling CTA to trigger follow-up and multi wavelength observations.  A system that automatically analyzes data coming from Cherenkov telescopes and generates alerts from CTA during the data acquisition is mandatory to  capture these phenomena during their evolution and for effective communication to the astrophysical community: this will be accomplished by means of an On-Site Analysis, in which a Real-Time Analysis pipeline is present, becoming a key system of the CTA observatory for the follow-up study of multi-wavelength and multi-messenger transient sources.

\section{The On-Site Analysis}

The CTA On-Site Analysis is a hardware and software system for the  analysis of scientific data of the Cherenkov events and for the instrument performance/data quality monitoring of the cameras that will be performed on the same site as the telescopes. 

CTA is expected to generate a large data rate, due to the large number of telescopes, the amount of pixels in each telescope camera and the fact that several time samples are recorded in each pixel of the triggered telescopes. Depending on (i) the array layout, (ii) the data reduction schema adopted, and (iii) the trigger criteria, the data rate estimations vary from 0.5 to 8 GiB/s. The location of the two arrays, where it is reasonable to suppose that the  sites will have a limited bandwidth for data transfer from the array site (North and South) to the Science Data Center (that receives the data from the arrays) implies that the data transfer of the raw data could require some days.

Due to the high data rate from CTA arrays and due to the limited bandwidth between the telescope sites and the Science Data Center, the On-Site Analysis should be co-located with the telescopes as an essential component of the CTA on-site infrastructure.

The On-Site Analysis has two pipelines: (i) the Level A or Real-Time Analysis, and (ii) the Level B analysis. Both pipelines have a dual purpose of delivering science feedback from the CTA data and monitoring  the instrument performance and the data quality. 

The On-Site Analysis is connected with the Data Acquisition, Archive and Array Control systems. Both pipelines read raw data, apply  calibration algorithms, reconstruct Cherenkov events and produce event lists; the focus of the Level-A Analysis is to maximize the speed of the analysis and the focus of the Level-B Analysis is to maximize the sensitivity. The Level-B Analysis also produces intermediate level data products, in addition to the final event list.

The Real-Time Analysis must be capable of generating science alert during observations or provide real-time feedbacks to external triggered ToO events with a latency of 30 s starting from the last acquired event that contributes to the alert. The Real-Time Analysis  differential sensitivity should be as close as possible to the one of the final offline analysis pipeline but not worse than a factor of 3: despite those caveats, an unprecedented sensitivity for short term exposures is achieved \cite{bib:Fioretti}.  The search for transient phenomena must be performed on multiple timescales (i.e. using different integration time windows) from seconds to hours, both within an a-priori defined source region, and elsewhere in the FoV. A Real-Time Analysis pipeline specific to each sub-array configuration is foreseen to run in parallel. The availability of the Real-Time analysis during observations must be greater than 98\%. 

The Real-Time Analysis enables CTA to provide real-time feedback to external received alerts, and to maximize the science return on time-variable and transient phenomena by issuing alerts of unexpected events from astrophysical sources in case of (i) Gamma-Ray Bursts (GRB) \cite{bib:Inoue}; (ii) serendipitous discoveries in the Field of View of the array, e.g. during the Galactic and extra-Galactic surveys; (iii) detecting particularly interesting states about a given source under observation. For this kind of CTA self-trigged alerts the main purpose of these science alerts is to re-schedule the observations to follow the phenomena in real-time and to issue alerts to other facilities.

The Level-B Analysis should be able to provide science feedback with a greater sensitivity with respect to the Real-Time Analysis, as close as possible to the final sensitivity of the CTA but with an integral sensitivity not worse than a factor of 2. The required maximum latency of the Level-B Analysis is 10 hours: this means that the Level-B Analysis should run at the end of each observation (if there will be enough computing power on-site) or next morning, when the telescopes and the Real-Time Analysis go off-line.

The Level-A should perform instrument performance and data quality tasks during the data acquisition;  the focus is the speed of the analysis to identify problems in the shortest possible time and notify them in real-time to the Array Control system. The Level-B Analysis should perform instrument performance and data quality monitoring with more details and  analyzing also historical data.

The development  of the On-Site Analysis system involves two Work Packages of the current CTA Product Breakdown Structure: the Data Management (DATA) \cite{bib:Lamanna} that has the responsibility of the On-Site Analysis software development, and the Array Control and Data Acquisition (ACTL) \cite{bib:Oya}, that has the responsibility to provide the on-site Information and Communication Technology (ICT) infrastructure (networking/computing architecture, cost estimations) to allow the On-Site Analysis to successfully run.

A deeper study for the definition of detailed use cases of the On-Site Analysis (with a special focus on the Real-Time Analysis) is in progress with the involvement of people from Data Management, Array Control and scientists expert on different fields. The  starting point are the CTA science cases and the main purpose is the definition of the steps of interaction with the CTA system to maximize the science return of each observation in the management of external triggers (by other observatories/facilities) or self-triggered by the Real-Time Analysis. 

\subsection{Architecture}

\begin{figure}
     \includegraphics[width=1\textwidth]{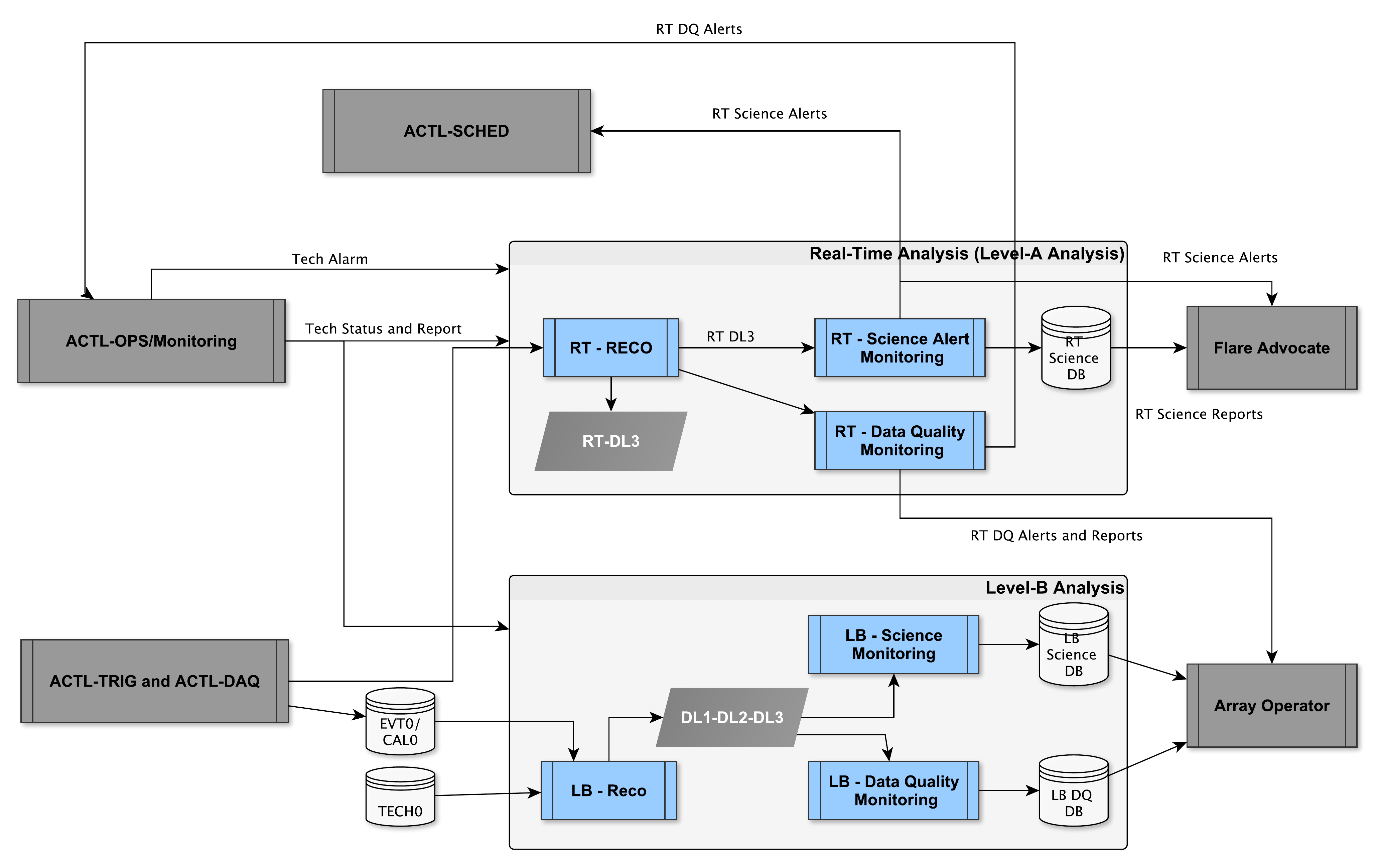}
     \caption{The On-Site Analysis Architecture.}
     \label{fig1}
     \end{figure}

A general view of the OSA architecture is reported in Figure \ref{fig1}. On-Site Analysis manages the  (i) Camera  data (namely both EVT0, i.e. events from Cherenkov cameras, and CAL0, i.e. calibration runs), and the (ii) Technical data TECH0 (data from auxiliary devices of the array and house keeping from cameras and telescope systems).

The On-Site Analysis interfaces with the (1) ACTL Data Acquisition system, in charge of both buffering camera data in the local repository and delivering them in streaming to the Real-Time Analysis, (2)  ACTL Scheduler, in particular the short-term scheduler system which receives and manages the science alerts generated by the Real-Time Analysis, (3)  ACTL Monitoring and Control system, in charge of the health and performance monitoring of any CTA assembly.

The Real-Time Analysis pipeline has the following components:

\begin{enumerate}
\item  the Real Time (RT) Reconstruction components, that perform a fast reconstruction of the acquired events;
\item  the RT Science Alert Monitoring components, that detect unexpected astrophysical events and generate Science Alerts;
\item  the RT Data Quality (DQ) Monitoring components, that perform a basic data quality check to evaluate the correct execution of the observations in real-time. This system generates RT DQ Alerts and Reports (statistics, graphs, and summaries of the data generated by the real-time analysis).
\end{enumerate}

This pipeline receives the technical alarm notification delivered by the ACTL system (TECH Alarms). TECH Alarms are urgent notifications given when data fall outside pre-defined quality criteria, signifying a failure in the system.  The Real-Time Analysis scientific results (the RT Science Alerts) should be stored in a database (the RT Science DB) with the produced event list (RT DL3 Repo). The real-time scientific reports (RT Science Report), which are the summary of the analysis results from the data being analyzed by the Real-Time Analysis will be retrieved  from a real-time dedicated DB (RT Science DB). 

The Level B analysis pipeline has the following modules:
\begin{enumerate}
\item The Level-B Data Quality (DQ) Monitoring performs a detailed data quality check to evaluate the correct execution of the observations. These results are stored into a dedicated data base (LB DQ DB), used to generate dedicated data quality reports (LB DQ Reports).
\item The Science Monitoring can analyze the data produced by the reconstruction software which runs within the Level-B Analysis pipeline (LB-RECO) to check the detection and the status of astrophysical sources with a sensitivity that is expected to be closer to the offline ones and certainly improved compared to the Level A analysis. It generates Science Reports.
\end{enumerate}

It is foreseen that the Level-B analysis will be the same as the off-site analysis pipeline (also called Level-C pipeline) sharing the same work. 

To fulfill the RTA requirements, taking into account the constraints in terms of data rate we have designed the RTA as an on-line system (the data are analyzed during the data acquisition) that acquires data from the Data Acquisition (DAQ) system of the CTA in streaming, reducing at the minimum level the overall latency of the system. More details are reported in \cite{bib:Bulgarelli2}. 

\subsection{Prototyping activities}

Some prototyping activities and software development of the analysis pipelines has been done by the CTA Consortium during the last two years, to explore some of the design ideas.  More details in the prototyping activities are reported in \cite{bib:Lamanna}. In the following we focus on the prototyping activities for the Real-Time Analysis pipeline.

An RTA Prototype was created to help understand how to fulfill the CTA requirements and  help to define  the Real-Time Analysis specifications. The  aim is to prove the feasibility of this analysis, taking into account the  mentioned constraints and requirements. The main purposes of the prototype is: (i) to prove the feasibility of the stream processing performed on CTA data; (ii) to test a fast inter-process communication architecture, focusing on both processes running on the same computing node (in-memory data transfer) and in the point-to-point data transfer between nodes; (iii) to test the feasibility of the stream processing with FPGA and GPU hardware accelerators.

For the development of this technological assessment, a simulator of the CTA Event Builder has been developed using as input the CTA Monte Carlo (MC) 'PROD2' data\cite{bib:bern2013}. The MC data are converted into a raw data format, a stream of bytes that are transmitted between the different processes of the RTA pipeline. Some reconstruction algorithms have been developed, in particular a waveform extraction algorithm\cite{bib:albert}, one of the most data throughput intensive algorithms, cleaning and Hillas parameter extraction. A first test has been performed to test inter-process communication on the same machine (an Intel Core i7 2.6 GHz, 4 cores, 16 GB 1600 MHz DDR3 RAM) and the data transfer is performed only in memory (no network data transfer and disk access are tested): the sustained data rate is between 2 and 2.4 GB/s, equivalent to an array trigger rate of about 4 kHz with the camera data with full waveforms. A consistent increment of the prototype performance is expected adding more CPUs.

The RTA prototype has been connected with the DAQ prototype\cite{bib:lyard} to provide the full chain of data acquisition and processing for CTA; data structures are serialized and reflected using Google Protocol Buffers, and a method is provided to write them efficiently to FITS tables, including complex zero-suppressed data. Visualization of data structures are made using OpenGL and a simple GUI framework. 

To take into account hardware accelerators we have tested the data transfer between CPU and GPU and the GPU speed, testing the waveform extraction algorithm \cite{bib:Rodriguez}. For the tests we have taken into account a camera with 2,048 pixels, 10 kHz, 50 samples per waveform and 2 bytes per sample. From our test we got 5 GB/s on PCIe 2.0 cards and 7 GB/s with PCIe 3.0 K40.

We have performed tests with an FPGA mounted on an IBM Power System S824 server to test the on-the-fly compression of the data (because "ready-to-use"). This test is useful to understand how the use of FPGAs could help to increase the final data rate of the overall RTA pipeline.  IBM Research Labs had performed some tests reaching 2 GB/s of compression rate with a compression ratio of 2.5 using raw format MC data (842 compression algorithm), using not more than 1.5\% of the area of the device's resources, leaving the rest for other functions (e.g. some RECO steps). It is possible to use multiple compression engines on the same FPGA to increase the bandwidth of the overall system. Based only on this first test we can save about 30 CPU cores, energy and network infrastructure with a fraction of a single FPGA.

We have used OpenCL for the development of CTA algorithms using hardware accelerators and CPUs. See  \cite{bib:Zoli} for more details.

\section{The Data Volume Reduction pipeline}

Into the current design schema  a Data Volume Reduction software is foreseen with the main purpose of reducing the volume of CTA data to an acceptable level without significant negative impact on performances. The three types of data-reduction  are under discussion:
\begin{enumerate}
\item waveform extraction: most of the CTA cameras record signal as a function of time in each pixel, but in general  only the signal amplitude and possibly arrival time are used in higher level analysis. After the commissioning phase only the 2\% of the pixels will be kept with the full waveform information;
\item zero-suppression: for the wide field-of-view cameras of CTA it will often be the case that only a few percent of all pixels contain useful information in a given event. Pixels which are not useful for image analysis can be discarded early to save on later processing and storage;
\item background rejection: the vast majority of CTA events will be proton or helium nuclei initiated cascades.  Basic event reconstruction in combination with some basic cuts could be used to discard events that are clearly background.
\end{enumerate}

To perform these tasks it will be possible to use some steps of the Real-Time Analysis pipeline as a pre-storage analysis for event pre-selection  within the Data Acquisition System for data reduction purposes.

\section{Conclusion}

The On-Site Analysis will perform a first CTA scientific analysis of data acquired from the array of telescopes, in both northern and southern sites. The latency of the system, in particular for the Real-Time Analysis and the sensitivity requirements are challenging because of the large data rate, a few GByte/s, and the complexity of the data analysis. Due to the high data rate and  the limited bandwidth between the telescope sites and the Science Data Center, the On-Site Analysis will be co-located with the telescopes as an essential component of the CTA on-site infrastructure. Substantial efforts  are envisaged to enable efficient data processing, taking into account the limited electrical and computer power available on-site. The performed prototyping activities prove the feasibility  of the Real-Time Analysis within the requirements imposed by CTA. 

To maximize the science return of each observation and the management of external or internal scientific alerts the definition of detailed use cases of the On-Site Analysis, linked with detailed studies on sensitivity, is in progress.

Thanks to the unprecedented sensitivity that the Real-Time Analysis will be able to achieve, that will be improved also by the technological studies  and by the optimization of data analysis algorithms that are in progress, CTA will be capable to follow the astrophysical phenomena in real-time. This will provide both real-time feedback to external received alerts from multi-wavelength and multi-messenger campaigns, and to issue alerts to other facilities, enabling CTA to trigger follow-up and multi wavelength observations. This will enable CTA to maximize the science return on time-variable and transient phenomena in a multi-wavelength and multi-messenger context.

\acknowledgments
We gratefully acknowledge support from the agencies and organizations 
listed under Funding Agencies at this website: http://www.cta-observatory.org/. We gratefully acknowledge support from IBM Labs to test compression techniques.

\end{document}